\documentclass[prl,twocolumn,showpacs,superscriptaddress]{revtex4}
\newcommand{\pd}[2]{\frac{\partial #2}{\partial #1}}

\usepackage{epsfig}
\begin{document}

\title{Multiplicative Noise Induces Zero Critical Frequency}

\author{I. Peleg}
%\affiliation{Department of Physics, Bar Ilan University, Ramat-Gan
%52900, Israel}
\author{E. Barkai}
\affiliation{Department of Physics, Bar Ilan University, Ramat-Gan
52900, Israel}

%02.50.-r Probability theory, stochastic processes, and statistics
%05.40.-a Fluctuation phenomena, random processes, noise, and Brownian motion

\begin{abstract}

 Stochastic Bloch equations
which model the fluorescence
of two level molecules and atoms,
NMR experiments and Josephson junctions are investigated to illustrate
the profound effect of multiplicative noise on the critical frequency of
a dynamical system. Using exact solutions and the cumulant expansion
we find two main effects: (i)
even very weak noise may double or triple the
number of critical frequencies,
 which is related to an instability of the system and (ii)
strong multiplicative noise may induce a non-trivial zero
critical frequency thus wiping out the over-damped phase.
\end{abstract}

\maketitle

 Many dissipative deterministic
dynamical systems exhibit two phases of motion:
an under-damped oscillatory
behavior or an over-damped non-oscillatory motion. The transition between
these common behaviors defines the critical frequency of the system,
e.g. the critical frequency of the damped harmonic oscillator.
Multiplicative noise is known to influence deterministic
systems, in profound and surprising ways
\cite{Gitterman,Lindenberg,Nonequalibrium Phase Transition 1, Stochastic Resonance 1, Gitterman1, Stochastic Resonance 2, Power Law 1, Power Law 2}.
Here we show how a stochastic perturbation induces
a zero critical frequency
for a particular non-trivial
choice of noise strength,
thus completely wiping out the over-damped phase.
 This might be  counter intuitive
at first glance, since
we expect noise to work against oscillations, however as we soon demonstrate,
in some cases the opposite situation is found. The second interesting
result we obtain is that even weak multiplicative noise may induce a
doubling or a tripling of the number of critical frequencies of a
system (in a way defined later)
a result which is related to an instability of the noiseless
dynamical system. Our results show how multiplicative noise
may influence the critical frequency of a system in profound ways.

 We investigate the dynamics of the
stochastic Bloch equation. The Bloch equation finds its
applications, in many fields of Physics ranging
from Nuclear Magnetic Resonance (NMR) \cite{NMR - Bloch Equations 1,remark}
to single molecule
spectroscopy \cite{SMS - Bloch Equations 1,Line Shape} and Josephson's junctions \cite{josephson 1}.
We use the example of the optical Bloch equation, however
with minor modifications we may consider other systems e.g. magnetic
systems. In particular we consider a two level electronic transition of an
atom or a molecule,
interacting with a continuous wave laser and a stochastic bath.
The optical Bloch equation for
$\vec{Z}(t) = (u,v,w,y)$, where $(u,v,w)$ describes the usual Bloch vector, is
$$ \frac{d}{dt}\vec{Z}(t)=M(t)\vec{Z}(t)  $$
\begin{equation}
\label{Bloch eq matrix}
 \quad M(t)= \left (
\begin{array}{cccc}
 -\frac{\Gamma}{2}  & \delta _L(t)             & 0       & 0\\
 -\delta _L(t)      & -\frac{\Gamma}{2}     & -\Omega    & 0 \\
 0                  & \Omega                & -\Gamma    & -\Gamma\\
 0                  & 0                     & 0          & 0
\end{array}
\right ).
\end{equation}
The initial condition is $\vec{Z}(0)=(0,0,-1/2,1/2)$
describing a system in the ground state and $y=1/2$ for all times.
Here $\Gamma$ is the radiate emission rate and $\Omega$ is the Rabi frequency
describing the interaction of the transition dipole of the system
with the laser field (within rotating wave approximation).
The stochastic detuning  $\delta_L(t)  =\omega_L-(\omega_0+\nu(t))$
describes the interaction of the system with the bath
in the spirit of the Kubo-Anderson line shape theory
\cite{SMS - Bloch Equations 1,Tanimura,Kubo},
namely $\delta_L(t)$ is
a stochastic process describing spectral diffusion.
$\omega_L$ is the laser frequency and $\omega_0$ is the absorption frequency
of the two level system.
Spectral diffusion is found in many
molecular, atomic and magnetic systems
and is well investigated \cite{SMS - Bloch Equations 1,Mukamel}.
The noise is called multiplicative since in Eq.
(\ref{Bloch eq matrix})
the spectral diffusion process multiplies the vector $\vec{Z}$.
Beyond spectral diffusion the equations describe a two level
system in the process of resonance fluorescence, where the laser frequency
exhibits fluctuations, namely the detuning is a random function of time.

For the noiseless case $\nu(t)=0$
and for zero laser detuning $\omega_L = \omega_0$
we have a
simple damped harmonic oscillator for $w$
\begin{equation}
\ddot{w}+\left(\frac{3\Gamma}{2} \right)\dot{w}+\left(\Omega^2+\frac{\Gamma^2}{2} \right)w+\left(\frac{\Gamma^2}{4} \right)=0.
\label{eqho}
\end{equation}
When $\Omega$ is larger than the critical frequency
$\Omega_c=\Gamma/4$
the system exhibits under-damped Rabi oscillations
while when $\Omega<\Omega_c$ it decays to
the steady state monotonically.
The critical frequency provides the quickest approach
of  the amplitude of the
damped harmonic oscillator to zero.

Now we consider
the system in the presence of the spectral noise and investigate
the average behavior $\langle w \rangle$. What will happen to the critical
frequency of the system? and can we choose parameters of the noise in such
a way that the critical frequency of the noisy  system is zero?

The formal solution  of the problem is
given in terms of the time ordered exponential
\begin{equation}
\label{formal solution}
\langle \vec{Z}(t) \rangle=
\left< \hat{T}\left\{\exp\left(\int_{0}^{t} M(\tau) d\tau \right)\right\} \right> \vec{Z}(0).
\end{equation}
In practice  it is generally difficult to
find explicit solutions due to
the combination of the time ordering operator $\hat{T}$ and the average
over the multiplicative stochastic process denoted
with $\langle \cdots \rangle$ in Eq.
(\ref{formal solution}). Here we find an exact solution
for a dichotomic two state Kubo-Anderson process \cite{Kubo}.
With this solution we will
explore whether the motion is over-damped or under-damped.
We later show that our findings are general beyond the exactly solvable
two state process.
In particular we consider $\nu(t)=\nu h(t)$, where $h(t)=+1$ or $h(t)=-1$
describes the stochastic two state
process with a rate $R$ for transitions between $+1$ and $-1$.
Such a model is applicable in single molecule spectroscopy
in glasses \cite{Glasses 1,Glasses 2,Line Shape} and was used extensively
for the line shape theory of Kubo and Anderson.

\begin{figure}
\begin{center}
\epsfxsize=80mm
\epsfbox{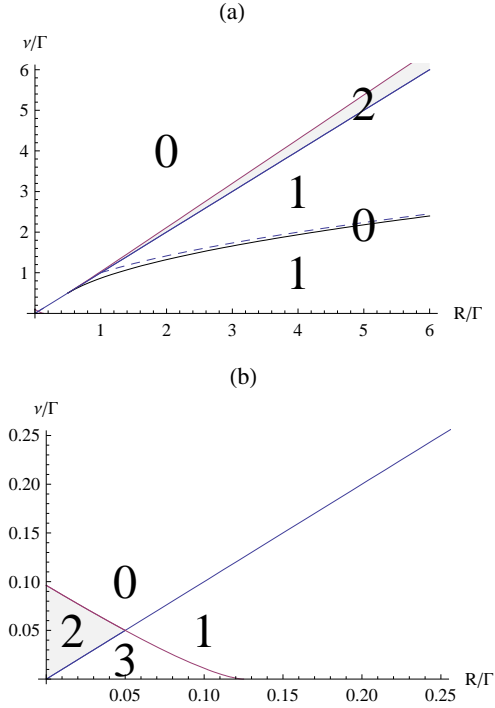}
\end{center}
\caption{
The phase diagram for the critical frequencies of the optical Bloch equation
with multiplicative two state noise. The $0,1,2,3$ indicate the number
of  critical frequencies, as defined in the text. In (a) we see
a line of zero critical frequency in the fast modulation limit $\nu< R$,
which is well approximated by the cumulant expansion
(the dashed
line
$\nu/\Gamma= \sqrt{ R/\Gamma}$).
In the absence of spectral diffusion $\nu=0$ we have a single critical
point, thus as shown in (b) the addition of weak noise $\nu/\Gamma<<1$
may strongly influence the critical frequency in the sense that
we find there a phase with $2$ or $3$ critical frequencies.
}
\label{fig1}
\end{figure}

We use Burshtein's method
\cite{Burshtein} of marginal averages to solve the Kubo-Anderson process
with zero laser  detuning $\omega_L = \omega_0$. We define the marginal
average vector
$(\langle\vec{Z}(t) \rangle_{+} , \langle\vec{Z}(t)\rangle_{-})$
which is the average of $\vec{Z}(t)$ given that at time
$t$ the stochastic process had the value $\pm1$ correspondingly. The
equation of motion for the marginal average vector is
\begin{equation}
    \pd{t}{}
    \left(
    \begin{array}{c}
        \langle\vec{Z}\rangle_+  \\
        \langle\vec{Z}\rangle_-
    \end{array}
    \right)
    =
    \left(
    \begin{array}{cc}
        A_0^{+} - R I & R I \\
        R I    & A_0^{-} - R I
    \end{array}
    \right)
    \left(
    \begin{array}{c}
        \langle\vec {Z}\rangle_+  \\
        \langle\vec {Z}\rangle_-
    \end{array}
    \right)
\label{eq03}
\end{equation}
\begin{equation}
A_0^{\pm} = \left(
\begin{array}{llll}
 -\frac{\Gamma }{2} & \pm\nu & 0 & 0 \\
 \mp\nu & -\frac{\Gamma }{2} & -\Omega  & 0 \\
 0 & \Omega  & -\Gamma  & -\Gamma  \\
 0 & 0 & 0 & 0
\end{array}
\right)
\label{A0}
\end{equation}
where $I$ is the identity matrix.
The operators  $A_{0}^{+}$ and $A_{0}^{-}$ are Bloch matrices
corresponding to
the state of the spectral diffusion
$\delta_L=+\nu$ or $\delta_L = - \nu$ respectively.
To solve the problem we must diagonalize the $8 \times 8$ matrix
in Eq. (\ref{eq03}), then complex eigenvalues yield under-damped oscillatory
modes while real eigenvalues correspond to over-damped modes.
The eigenvalues $\{ \lambda \}$ are found using the
characteristic polynomial of
Eq. (\ref{eq03})
\begin{equation}
\lambda(\lambda+2R)P_1(\lambda)P_2(\lambda)=0,
\label{factors}
\end{equation}
where two cubic polynomials are defined as
%
%\begin{widetext}
%
\begin{equation}
\label{sd polynomial}
\begin{array}{rl}
P_1(\lambda)=   &\left(8\Omega ^2+4 (\Gamma +\lambda ) (\Gamma +2 \lambda )\right)R + \\
                &+2(\Gamma +2 \lambda ) \Omega ^2+(\Gamma +\lambda ) \left((\Gamma +2 \lambda )^2+4 \nu ^2\right)\\
\\
P_2(\lambda)=   &8(\Gamma +2 \lambda )R^2 + \\
                &+2 \left(4 \nu ^2+(\Gamma +2 \lambda ) (3 \Gamma +4 \lambda )\right)R + \\
                &+2(\Gamma +2\lambda ) \Omega ^2+(\Gamma +\lambda ) \left((\Gamma +2 \lambda )^2+4 \nu ^2\right).
\end{array}
\end{equation}
We have thus reduced the problem to finding the roots of two
third order  polynomials
$P_1(\lambda) = 0$ and $P_2(\lambda) = 0$.
After some algebra one can show that only
the roots of $P_1(\lambda)$ enter the solution of
$\langle \vec{Z} \rangle = \langle \vec{Z} \rangle_{+} + \langle \vec{Z} \rangle_{-}$
as well as the eigenvalue  $\lambda=0$
[see Eq. (\ref{factors})] which
yields the
steady state solution \cite{remark2}.

 A physical observable is the intensity of emitted light
$\langle \mathcal{I} \left( t \right) \rangle$ which is equal
to $\Gamma$ times the population in the excited state
$\langle \mathcal{I} \left( t \right) \rangle \equiv
\Gamma \left( \langle w \rangle + {1 \over 2} \right)$.
The eigenvalues  $\{ \lambda_1, \lambda_2, \lambda_3 \}$
are the solutions of $P_1(\lambda)= 0$ and the intensity
is
\begin{equation}
\begin{array}{l}
\langle \mathcal{I}(t) \rangle = {\cal I}_{ss} -
\\\\
-\sum_{i = 1 } ^3 e^{t \lambda _n}
\frac{\Gamma  \left(\Gamma +\lambda _n\right)
   \left(4 R \Gamma _{\text{SD}}+\left(\Gamma +2 \lambda _n\right)
   \left(4 R+\Gamma +2 \lambda _n\right)\right)}
   {2 \lambda _n\left(5 \Gamma ^2+12 R \Gamma +4 \Omega ^2
   +4 R \Gamma_{\text{SD}}
   +4 \lambda _n \left(4 (R+\Gamma )+3 \lambda_n\right)\right)}
\end{array}
\end{equation}
with $\Gamma_\mathrm{SD} \equiv \nu^2 / R$.
The steady state solution is
\begin{equation}
\mathcal{I}_{ss} =
\frac{\Gamma  (4 R+\Gamma ) \Omega ^2}
{(4 R+\Gamma ) \left(\Gamma^2+2 \Omega ^2\right)+4 R \Gamma  \Gamma _{\text{SD}}}.
\end{equation}
This expression when $\Omega \to 0$  is the
well known Kubo-Anderson line shape at zero laser detuning.

We now focus our attention on the eigenvalues
$\{\lambda_i\}$ to determine whether the solution is
over-damped
or under-damped. The motion is called over-damped if
all eigenvalues  $\{ \lambda_i \}$ are real  otherwise it is
under-damped.
The condition for over-damped behavior is that the
discriminant $\mathcal{D}$
 of $P_1 (\lambda)$ be less than zero, explicitly we have
\begin{widetext}
\begin{equation}
\begin{array}{ll}
\mathcal{D} =
&
-16384 \left(\left(\frac{\Gamma -\Gamma _{\text{SD}}}{4}\right)^2-\Omega ^2\right) R^4
+ 512 \left(\Gamma ^3-8 \Omega ^2 \left(2 \Gamma +5 \Gamma
   _{\text{SD}}\right)+\Gamma _{\text{SD}} \left(\Gamma ^2+2 \Gamma
   _{\text{SD}} \left(\Gamma _{\text{SD}}-2 \Gamma
   \right)\right)\right) R^3
\\
&+64 \left(128 \Omega ^4+8 \left(\Gamma ^2+19 \Gamma _{\text{SD}}
   \Gamma +6 \Gamma _{\text{SD}}^2\right) \Omega ^2-\Gamma ^2
   \left(\Gamma ^2+8 \left(\Gamma -\Gamma _{\text{SD}}\right)
   \Gamma _{\text{SD}}\right)\right) R^2
\\
&+ 64 \left(\Gamma _{\text{SD}} \Gamma ^4+2 \Omega ^2 \left(\Gamma -10
   \Gamma _{\text{SD}}\right) \Gamma ^2+16 \Omega ^4 \left(3 \Gamma
   _{\text{SD}}-2 \Gamma \right)\right) R
-1024 \Omega ^4 \left (\left (\frac{\Gamma}{4}\right) ^2 - \Omega ^2 \right) < 0.
\end{array}
\label{SD Discriminant}
\end{equation}
\end{widetext}

Recall that for the noiseless case we have a single
critical frequency $\Omega_c = \Gamma/4$.
 As shown in Fig. \ref{fig1}
in the presence of multiplicative
noise the phase diagram of the motion is  very rich: \\
(i) In the slow modulation $\nu \gg R$ strong coupling $\nu/\Gamma \gg 1$
regime, the solutions are always oscillatory and the critical frequency is $0$
(indicated by $0$ in Fig \ref{fig1}a).\\
(ii) When $\nu \approx R$ we obtain
over-damped motion when
$\Omega_{C_1} < \Omega <\Omega_{C_2}$
so we have \emph{two} critical frequencies
(indicated by $2$ in Fig. \ref{fig1}a).\\
(iii) In the fast modulation  $R > \nu$ strong coupling $\nu /\Gamma>1$
limit,
we find a single critical frequency similar to the noiseless case,
except for a  surprising
line on which $\Omega_C=0$ (denoted with $0$ in
Fig. \ref{fig1}a).\\
(iv) In the weak noise limit $\nu \ll \Gamma$,
the solution yields either $2$ or  $3$ critical frequencies
(see Fig. \ref{fig1}b).
Thus weak noise modifies the solution dramatically by doubling
or tripling the number of critical frequencies of the system.

\begin{figure}
\begin{center}
\epsfxsize=80mm
\epsfbox{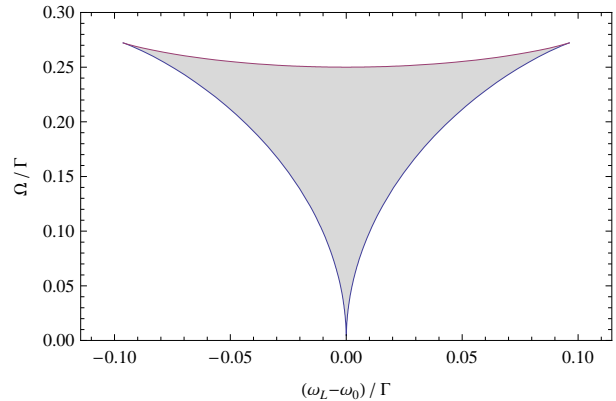}
\end{center}
\centering
\caption{Phase diagram of the optical Bloch equation in the absence
of the multiplicative noise.
The darker area is the over-damped phase.
For zero detuning $\omega_L-\omega_0=0$ the critical frequency is $\Omega_C=\Gamma/4$.
Notice the cusp at zero detuning which makes the solutions unstable
to multiplicative noise.
}
\label{fig2}
\end{figure}

 To understand better the phase diagram Fig. \ref{fig1}
we present in Fig. \ref{fig2} the behavior of the solution
of the optical Bloch equation in the {\em  absence} of the multiplicative
noise
i.e. $\nu=0$.
The figure is a phase diagram  in the Rabi frequency $\Omega/\Gamma$
and  detuning $(\omega_L- \omega)/\Gamma$ plane showing the regions of
over-damped and under-damped behavior.
Fig. \ref{fig2}
illustrates that the noiseless system
is unstable in the
sense that for any small detuning
and low enough Rabi frequency we get an
oscillatory behavior while for zero detuning the solution is
over-damped.
This instability of the noiseless solution explains why
even adding a weak  perturbation  $\nu\ll\Gamma$ strongly affects the system.
Namely, for weak noise (Fig, \ref{fig1}b) we find either $2$ or $3$
critical frequencies instead of 1 for the noiseless case.

 As mentioned before for slow modulation
$\nu > R$
and strong coupling
$\nu \gg \Gamma$
the motion is always under-damped.
To understand this behavior we again refer to
the noiseless case presented in Fig. \ref{fig2}, where we observe that
large detuning means an oscillatory solution.
Namely oscillations for the  noise free Bloch equation are induced by
two mechanisms, the Rabi frequency and the detuning.
Hence it is not surprising that strong and slow noise in Fig. \ref{fig1}a
(i.e. $\nu >R,\Gamma$)
may
induce oscillations and the wipe out of the over-damped motion.

 Far less trivial is the wipe out of over-damped motion
in the  fast modulation limit, i.e. the line
of zero critical frequency in Fig. \ref{fig1}a.
To investigate this behavior we consider the limit $R \to \infty$.
Then using Eq.
(\ref{SD Discriminant}) we find
the critical frequency
\begin{equation}
\label{SD Fast modulation critical frequency}
\lim_{\nu,R \to\infty}
\Omega_C=\left|\frac{\Gamma-\Gamma_\mathrm{SD}}{4}\right|
\label{eq9}
\end{equation}
where the limit is taken with $\Gamma_\mathrm{SD}$ remaining finite.
We see that $\Omega_c = 0$ when $\Gamma_\mathrm{SD} = \Gamma$,
namely when $\nu/\Gamma = \sqrt{R /\Gamma}$. This line is shown
in Fig. \ref{fig1} as a dashed line.

\begin{figure}
\begin{center}
\epsfxsize=80mm
\epsfbox{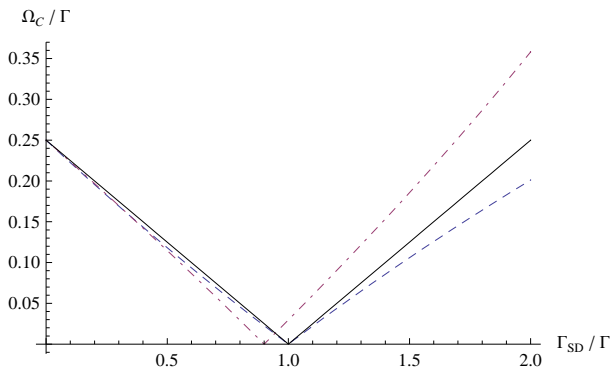}
\end{center}
\caption{ The critical frequency $\Omega_c$ as a
function of $\Gamma_\mathrm{SD}$ which is the measure of noise strength.
For small noise levels, i.e. $\Gamma_\mathrm{SD}/\Gamma<1$,
the critical frequency $\Omega_c$
decreases as anticipated in Eq.
(\ref{eq12}). The figure illustrates the existence of $\Omega_c=0$
for a particular noise value.
We show the  critical frequency obtained from the exact solution
(dot-dashed)
and the cumulant approximation (dashed) for $R=2.5
\Gamma$.
The solid line is the critical frequency $\Omega_c$ at $R\to\infty$
Eq. (\ref{eq9}).
}
\label{fig3}
\end{figure}

 Expanding  the exact solution in $\Gamma_\mathrm{SD}$, one can show that
for $R > \Gamma/8$ any amount of noise will lead to a decrease of
$\Omega_c$ according to
\begin{equation}
\label{eq12}
\Omega_C=\frac{\Gamma }{4}-\frac{2 \Gamma_\mathrm{SD}}{8  - \Gamma/R }+O\left(\Gamma_\mathrm{SD} ^2\right),
\end{equation}
where the leading $\Gamma/4$ term describes the noiseless case.
The decrease of $\Omega_c$ is explained by the fact that the noise removes the
system from  zero detuning and hence solutions
tend to be more oscillatory (i.e. the critical frequency is reduced).
The surprising result is that by  increasing the noise level we
reach a limit where the critical frequency is zero.
Such a behavior in the fast modulation limit could not be anticipated
without our mathematical analysis.
The behavior of $\Omega_c$ is illustrated in Fig. \ref{fig3},
which shows the decrease of the critical frequency until it reaches the
value $\Omega_c=0$.

 It is natural to ask if the behavior we found is general
or limited to the example of a two state process. For this
aim we have used the cumulant expansion \cite{Van Kampen}, to investigate the
critical frequency $\Omega_c$ of the system. We consider a
stationary process $h(t)$
whose correlation function is
$\langle  h(t) h(t + \tau) \rangle = \exp ( - R \tau)$.
The cumulant expansion works well when the Kubo number $\nu/R$ is small.
Within this approximation \cite{Van Kampen}
\begin{equation}
\label{cumulant equation} \pd{t}{\langle\vec Z\rangle}=
\left(A_0 +\nu^2 K\right)\langle\vec Z\rangle
\end{equation}
where $A_0 = A_0 ^{\pm}|_{\nu=0}$ and
\begin{equation}
\label{K}
K=\int_0^\infty e^{ - R \tau} A_1 e^{A_0 \tau} A_1
e^{-A_0\tau} d\tau
\end{equation}
with
$ A_1= \left(
\begin{array}{ll}
\sigma & 0 \\
0 &  0 \\
\end{array}
\right)$
and $
 \sigma= \left(
\begin{array}{ll}
0 & 1 \\
-1 &  0 \\
\end{array}
\right)$.
%
%$$
%A_1= \left(
%\begin{array}{llll}
% 0 & 1 & 0 & 0 \\
% -1& 0 & 0 & 0 \\
% 0 & 0 & 0 & 0 \\
% 0 & 0 & 0 & 0
%\end{array}
%\right).
%$$
Solving the integrals leads to cumbersome equations
for $\Omega_c$. However in the limit $R \to \infty$
and $\nu \to \infty$ in such a way that $\Gamma_\mathrm{SD}=\nu^2/R$ remains constant
we find that
Eq. (\ref{eq9}) is valid
and therefore the equation is not limited to the two state Kubo-Andersen model.
This means that for a large class of stochastic processes,
 multiplicative noise
induces zero critical frequency in the fast modulation limit and
the curve $\nu/\Gamma= \sqrt{ R/\Gamma}$, on which $\Omega_c=0$
shown
in Fig. \ref{fig1}(a),  is a general behavior.

To further validate the generality of our results we have solved
semi-analytically and with the help of Mathematica:
(i) two state model with two  non identical rates describing
the transitions between up and down states, and (ii) models with three
states. These models show behaviors similar to our findings.

The same  effects cannot be found for linear systems driven by additive
noise with zero mean, since the averaged equations have the same
critical frequency as the noiseless case. Hence for linear systems the
effects we have found are limited to systems with multiplicative noise.
However nonlinear systems with additive noise may exhibit behaviors identical
to those investigated in this manuscript.
 To see this we add a coordinate $\delta$ to the description of the system thus Eq. (\ref{Bloch eq matrix}) is written as:
\begin{equation}
\pd{t}{}
\left(
\begin{array}{c}
\delta \\
\vec{Z}
\end{array}
\right)
=
\left(
\begin{array}{cc}
0 & 0   \\
0 & M(\delta)
\end{array}
\right)
\left(
\begin{array}{c}
\delta \\
\vec{Z}
\end{array}
\right)
+
\left(
\begin{array}{c}
\xi(t)\\
0
\end{array}
\right)
.
\label{non linear}
\end{equation}
This equation is a non linear stochastic equation
with additive noise (i.e. $\xi(t)$ in Eq.
(\ref{non linear})) which is equivalent to the multiplicative Eq.
 (\ref{Bloch eq matrix}). We see that the two main
effects found in this manuscript: (a)
noise inducing zero critical frequency
 and
(b) the doubling or the tripling of the number of critical frequencies,
{\em even for weak noise},
may be found either in linear multiplicative systems or non
linear additive systems.
Thus we expect the main features of  our results to be valid for a vast class
of stochastic dynamical systems.

\textbf{Acknowledgment}
This work was supported by the Israel Science Foundation.


\begin{thebibliography}{99}
\bibitem{remark}
In NMR but also for optical systems
 additional relaxations described by $T_1$ and $T_2$ may be included
in the theory Eq.
(\ref{Bloch eq matrix}),
 however we will focus on the simplest case here.
\bibitem{remark2} Note that the roots of $P_2(\lambda)=0$ enter in the
solutions
of $\langle \vec{Z} \rangle_{+} - \langle \vec{Z} \rangle_{-}$.
\end{thebibliography}
\end{document}